\documentclass{article}

\usepackage[utf8]{inputenc}
\usepackage[english]{babel}
\usepackage{csquotes}

\newcommand{\ourTool}{\textbf{\texttt{MoULDyS}}}

\usepackage{subcaption}

\usepackage{paralist} %

\newenvironment{ienumerate}
	{\ifdefined\VersionLong\begin{enumerate}\else\begin{inparaenum}[\itshape i\upshape)]\fi}
	{\ifdefined\VersionLong\end{enumerate}\else\end{inparaenum}\fi}

\newenvironment{oneenumerate}
	{\ifdefined\VersionLong\begin{enumerate}\else\begin{inparaenum}[1)]\fi}
	{\ifdefined\VersionLong\end{enumerate}\else\end{inparaenum}\fi}

	\usepackage[backend=biber,backref=true,style=alphabetic,url=false,doi=true,defernumbers=true,sorting=anyt,maxnames=99]{biblatex} %
	\renewbibmacro*{doi+eprint+url}{%
		\iftoggle{bbx:doi}
			{\color{black!40}\footnotesize\printfield{doi}}
			{}%
		\newunit\newblock
		\iftoggle{bbx:eprint}
			{\usebibmacro{eprint}}
			{}%
		\newunit\newblock
		\iftoggle{bbx:url}
			{\usebibmacro{url+urldate}}
			{}%
	}

\ifdefined\VersionWithComments
	\usepackage{draftwatermark}
	\SetWatermarkText{draft}
	\SetWatermarkScale{15}
	\SetWatermarkColor[gray]{0.9}
\fi
\usepackage[svgnames,table]{xcolor}
\definecolor{USPNcobalt}{HTML}{293358}
\definecolor{USPNocre}{HTML}{8b7d6d}
\definecolor{USPNblanc}{HTML}{ffffff}
\definecolor{USPNceruleen}{HTML}{354878}
\definecolor{USPNsable}{HTML}{ad947e}

\usepackage[
		pdfauthor={Bineet Ghosh and Etienne Andre},%
		pdftitle={MoULDyS: Monitoring of Autonomous Systems in the Presence of Uncertainties},
		breaklinks  = true,
		colorlinks  = true,
	\ifdefined \VersionWithComments
	\fi
		citecolor   = USPNsable,
		linkcolor   = USPNocre,
		urlcolor    = USPNceruleen,
	]{hyperref}

\usepackage[capitalise,english,nameinlink]{cleveref} %
\usepackage{amssymb}
\newcommand{\imitator}{\textsf{IMITATOR}}

\ifdefined \VersionWithComments
 	\definecolor{colorok}{RGB}{80,80,150}
\else
	\definecolor{colorok}{RGB}{0,0,0}
\fi

\usepackage{xspace}

\newcommand{\eg}{\textcolor{colorok}{e.g.,}\xspace}

\newcommand{\ie}{\textcolor{colorok}{i.e.,}\xspace}

\newcommand{\license}{\texttt{gpl-3.0}}

\usepackage{tikz}
\usetikzlibrary{arrows,automata,calc,positioning} %

\tikzstyle{sample}=[color=blue,radius=7pt]
\tikzstyle{missingsample}=[color=blue!20,radius=4pt]
\tikzstyle{extrasample}=[color=green,radius=4pt]
\tikzstyle{signal}=[color=blue,-,densely dotted]
\tikzstyle{nodraw}=[draw=none,inner sep=0pt,minimum size=0pt]
\tikzstyle{uncertainsample}=[draw=blue,thick] %

\usepackage{subcaption}

\makeatletter
\def\orcidID#1{\smash{\href{https://orcid.org/#1}{\protect\raisebox{-1.25pt}{\protect\includegraphics{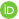}}}}}
\makeatother

\begin{document}

\title{\ourTool: Monitoring of Autonomous Systems in the Presence of Uncertainties\footnote{%
	This is the author version of the manuscript of the same name published in Science of Computer Programming.
	The final version is available at \href{https://doi.org/10.1016/j.scico.2023.102976}{\nolinkurl{10.1016/j.scico.2023.102976}}.
}}
\date{}

\maketitle

\noindent{}\textbf{Bineet Ghosh\orcidID{0000-0002-1371-2803}}
\\
{\em\small{}The University of North Carolina at Chapel Hill, NC, USA}

\smallskip

\noindent{}\textbf{Étienne André\orcidID{0000-0001-8473-9555}}
\\
{\em\small{}Université de Lorraine, CNRS, Inria, LORIA, F-54000 Nancy, France}
\\
{\em\small{}Université Sorbonne Paris Nord, LIPN, CNRS UMR 7030, F-93430 Villetaneuse, France}

\begin{abstract}
We introduce \ourTool{}, that implements efficient offline and online monitoring algorithms of black-box cyber-physical systems w.r.t.\ safety properties.
\ourTool{} takes as input an uncertain log (with noisy and missing samples), as well as a bounding model in the form of an uncertain linear system; this latter model plays the role of an over-approximation so as to reduce the number of false alarms.
\ourTool{} is Python-based and available under the \textit{GNU General Public License v3.0} (\license{}). %
We further provide easy-to-use scripts to recreate the results of two case studies introduced in an earlier work.

\end{abstract}

\noindent{}\textbf{Keywords:}
energy-aware monitoring, cyber-physical systems, formal methods, monitoring tool.

\section{Motivation and Significance}
\label{sec:background}

\begin{table}[!h]
\setlength{\tabcolsep}{2pt} %
	\scalebox{.87}{
	\begin{tabular}{|l|p{6.5cm}|p{6.5cm}|}
	\hline
	\textbf{Nr.} & \textbf{Code metadata description} & \textbf{} \\
	\hline
	C1 & Current code version & v1.1 \\
	\hline
	C2 & Permanent link to code/repository used for this code version &  {\scriptsize\url{https://github.com/bineet-coderep/MoULDyS/releases/tag/v1.1}} \\
	\hline
	C3  & Permanent link to Reproducible Capsule & \href{https://www.doi.org/10.5281/zenodo.7888502}{\nolinkurl{10.5281/zenodo.7888502}}\\
	\hline
	C4 & Legal Code License   & \textit{GNU General Public License v3.0} (\license{})~\cite{gpl} \\
	\hline
	C5 & Code versioning system used & \texttt{git} \\
	\hline
	C6 & Software code languages, tools, and services used & Python, \texttt{numpy}, \texttt{scipy}, \texttt{mpmath}, \texttt{pandas}, \texttt{Gurobi}\\
	\hline
	C7 & Compilation requirements, operating environments and dependencies & Provided in the installation guide~\cite{install_guide}.\\
	\hline
	C8 & If available, link to developer documentation/manual & Provided in the user guide~\cite{user_guide}.\\
	\hline
	C9 & Support email for questions & \texttt{bineet@cs.unc.edu}  %
		\\
	\hline
	\end{tabular}
	}
	\caption{Code metadata}
	\label{table:metadata}
\end{table}

Monitoring consists of analyzing system logs, \eg{} for detecting safety violations (see, \eg{} \cite{BDDFMNS18}).
Monitoring has many useful applications such as detecting the cause of a crash of vehicles.
As an example, autonomous systems are generally equipped with a device
that records their state at periodic or aperiodic time steps---logging the behavior of the system until the time of a failure.
A log, comprising of such recorded samples, is then investigated for possible safety violations.
Not only the logs can have samples missing at various time steps, but also the recorded samples can have added noise to it, \eg{} due to sensor uncertainties.
Analyzing such logs to detect possible safety violations, that might have caused a failure, is known as \textit{offline monitoring} when the analysis is done \emph{a posteriori} (see, \eg{} \cite{BCEHKM16}).
In contrast, it is \emph{online} when performed on-the-fly, when the whole log is not (yet) known (see \cite{Maler16} for a discussion on online verification).

We introduce here \ourTool{}\footnote{\url{https://sites.google.com/view/mouldys}} (see \cref{table:metadata} for code metadata), a monitoring tool to analyze logs to detect possible safety violations.
The specific features of \ourTool{} are twofold:
\begin{oneenumerate}
	\item the possibility to monitor aperiodic logs, or periodic logs with missing samples, and with possible noise over the recorded data; and
	\item the presence of a bounding model following the formalism of uncertain linear systems.
\end{oneenumerate}

\paragraph{Uncertain linear systems}
Uncertain linear systems~\cite{LP15,GD19} are a special subclass of non-linear systems that can be used to represent uncertainties and parameters in linear dynamical systems---for example, they can represent uncertain parameters in the cells of the dynamics matrix.
Such a formalism is useful in representing an over-approximation of the model when the precise model is unknown or hard to obtain.

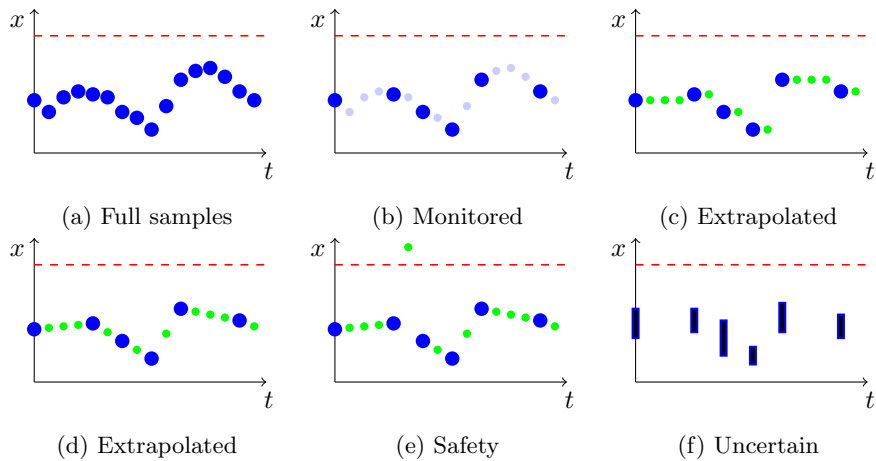
\begin{figure*}[tb]
	\newcommand{\imagewidth}{.32\linewidth}
	\begin{subfigure}[c]{\imagewidth}
		\begin{tikzpicture}[shorten >=1pt, scale=.65, yscale=.6, xscale=0.6, auto]

			\draw[->] (0, 0) --++ (0, 5.0) node[anchor=north east]{$x$};
			\draw[->] (0, 0) --++ (8.0, 0) node[anchor=north]{$t$};
			\draw[dashed, color=red, semithick] (0, 4) --++ (8.0, 0);

			\fill[sample] (0, 1.8) coordinate (s1) circle[];
			\fill[sample] (0.5, 1.4) coordinate (s2) circle[];
			\fill[sample] (1, 1.9) coordinate (s3) circle[];
			\fill[sample] (1.5, 2.1) coordinate (s4) circle[];
			\fill[sample] (2, 2) coordinate (s5) circle[];
			\fill[sample] (2.5, 1.9) coordinate (s6) circle[];
			\fill[sample] (3, 1.4) coordinate (s7) circle[];
			\fill[sample] (3.5, 1.2) coordinate (s8) circle[];
			\fill[sample] (4, 0.8) coordinate (s9) circle[];
			\fill[sample] (4.5, 1.6) coordinate (s10) circle[];
			\fill[sample] (5, 2.5) coordinate (s11) circle[];
			\fill[sample] (5.5, 2.8) coordinate (s12) circle[];
			\fill[sample] (6.0, 2.9) coordinate (s13) circle[];
			\fill[sample] (6.5, 2.6) coordinate (s14) circle[];
			\fill[sample] (7, 2.1) coordinate (s15) circle[];
			\fill[sample] (7.5, 1.8) coordinate (s16) circle[];

		\end{tikzpicture}
		\caption{Full samples}
		\label{example:full}
	\end{subfigure}
	\begin{subfigure}[c]{\imagewidth}
		\begin{tikzpicture}[shorten >=1pt, scale=.65, yscale=.6, xscale=0.6, auto]

			\draw[->] (0, 0) --++ (0, 5.0) node[anchor=north east]{$x$};
			\draw[->] (0, 0) --++ (8.0, 0) node[anchor=north]{$t$};
			\draw[dashed, color=red, semithick] (0, 4) --++ (8.0, 0);

			\fill[sample] (0, 1.8) coordinate (s1) circle[];
			\fill[missingsample] (0.5, 1.4) coordinate (s2) circle[];
			\fill[missingsample] (1, 1.9) coordinate (s3) circle[];
			\fill[missingsample] (1.5, 2.1) coordinate (s4) circle[];
			\fill[sample] (2, 2) coordinate (s5) circle[];
			\fill[missingsample] (2.5, 1.9) coordinate (s6) circle[];
			\fill[sample] (3, 1.4) coordinate (s7) circle[];
			\fill[missingsample] (3.5, 1.2) coordinate (s8) circle[];
			\fill[sample] (4, 0.8) coordinate (s9) circle[];
			\fill[missingsample] (4.5, 1.6) coordinate (s10) circle[];
			\fill[sample] (5, 2.5) coordinate (s11) circle[];
			\fill[missingsample] (5.5, 2.8) coordinate (s12) circle[];
			\fill[missingsample] (6.0, 2.9) coordinate (s13) circle[];
			\fill[missingsample] (6.5, 2.6) coordinate (s14) circle[];
			\fill[sample] (7, 2.1) coordinate (s15) circle[];
			\fill[missingsample] (7.5, 1.8) coordinate (s16) circle[];

		\end{tikzpicture}
		\caption{Monitored}
		\label{example:monitored}
	\end{subfigure}
	\begin{subfigure}[c]{\imagewidth}
		\begin{tikzpicture}[shorten >=1pt, scale=.65, yscale=.6, xscale=0.6, auto]

			\draw[->] (0, 0) --++ (0, 5.0) node[anchor=north east]{$x$};
			\draw[->] (0, 0) --++ (8.0, 0) node[anchor=north]{$t$};
			\draw[dashed, color=red, semithick] (0, 4) --++ (8.0, 0);

			\fill[sample] (0, 1.8) coordinate (s1) circle[];
			\fill[extrasample] (0.5, 1.8) coordinate (s2) circle[];
			\fill[extrasample] (1, 1.8) coordinate (s3) circle[];
			\fill[extrasample] (1.5, 1.8) coordinate (s4) circle[];
			\fill[sample] (2, 2) coordinate (s5) circle[];
			\fill[extrasample] (2.5, 2) coordinate (s6) circle[];
			\fill[sample] (3, 1.4) coordinate (s7) circle[];
			\fill[extrasample] (3.5, 1.4) coordinate (s8) circle[];
			\fill[sample] (4, 0.8) coordinate (s9) circle[];
			\fill[extrasample] (4.5, 0.8) coordinate (s10) circle[];
			\fill[sample] (5, 2.5) coordinate (s11) circle[];
			\fill[extrasample] (5.5, 2.5) coordinate (s12) circle[];
			\fill[extrasample] (6.0, 2.5) coordinate (s13) circle[];
			\fill[extrasample] (6.5, 2.5) coordinate (s14) circle[];
			\fill[sample] (7, 2.1) coordinate (s15) circle[];
			\fill[extrasample] (7.5, 2.1) coordinate (s16) circle[];

		\end{tikzpicture}
		\caption{Extrapolated}
		\label{example:linear}
	\end{subfigure}

	\begin{subfigure}[c]{\imagewidth}
		\begin{tikzpicture}[shorten >=1pt, scale=.65, yscale=.6, xscale=0.6, auto]

			\draw[->] (0, 0) --++ (0, 5.0) node[anchor=north east]{$x$};
			\draw[->] (0, 0) --++ (8.0, 0) node[anchor=north]{$t$};
			\draw[dashed, color=red, semithick] (0, 4) --++ (8.0, 0);

			\fill[sample] (0, 1.8) coordinate (s1) circle[];
			\fill[extrasample] (0.5, 1.85) coordinate (s2) circle[];
			\fill[extrasample] (1, 1.9) coordinate (s3) circle[];
			\fill[extrasample] (1.5, 1.95) coordinate (s4) circle[];
			\fill[sample] (2, 2) coordinate (s5) circle[];
			\fill[extrasample] (2.5, 1.7) coordinate (s6) circle[];
			\fill[sample] (3, 1.4) coordinate (s7) circle[];
			\fill[extrasample] (3.5, 1.1) coordinate (s8) circle[];
			\fill[sample] (4, 0.8) coordinate (s9) circle[];
			\fill[extrasample] (4.5, 3.3/2) coordinate (s10) circle[];
			\fill[sample] (5, 2.5) coordinate (s11) circle[];
			\fill[extrasample] (5.5, 2.4) coordinate (s12) circle[];
			\fill[extrasample] (6.0, 2.3) coordinate (s13) circle[];
			\fill[extrasample] (6.5, 2.2) coordinate (s14) circle[];
			\fill[sample] (7, 2.1) coordinate (s15) circle[];
			\fill[extrasample] (7.5, 1.9) coordinate (s16) circle[];

		\end{tikzpicture}
		\caption{Extrapolated}
		\label{example:constant}
	\end{subfigure}
	\begin{subfigure}[c]{\imagewidth}
		\begin{tikzpicture}[shorten >=1pt, scale=.65, yscale=.6, xscale=0.6, auto]

			\draw[->] (0, 0) --++ (0, 5.0) node[anchor=north east]{$x$};
			\draw[->] (0, 0) --++ (8.0, 0) node[anchor=north]{$t$};
			\draw[dashed, color=red, semithick] (0, 4) --++ (8.0, 0);

			\fill[sample] (0, 1.8) coordinate (s1) circle[];
			\fill[extrasample] (0.5, 1.85) coordinate (s2) circle[];
			\fill[extrasample] (1, 1.9) coordinate (s3) circle[];
			\fill[extrasample] (1.5, 1.95) coordinate (s4) circle[];
			\fill[sample] (2, 2) coordinate (s5) circle[];
			\fill[extrasample] (2.5, 4.6) coordinate (s6) circle[];
			\fill[sample] (3, 1.4) coordinate (s7) circle[];
			\fill[extrasample] (3.5, 1.1) coordinate (s8) circle[];
			\fill[sample] (4, 0.8) coordinate (s9) circle[];
			\fill[extrasample] (4.5, 3.3/2) coordinate (s10) circle[];
			\fill[sample] (5, 2.5) coordinate (s11) circle[];
			\fill[extrasample] (5.5, 2.4) coordinate (s12) circle[];
			\fill[extrasample] (6.0, 2.3) coordinate (s13) circle[];
			\fill[extrasample] (6.5, 2.2) coordinate (s14) circle[];
			\fill[sample] (7, 2.1) coordinate (s15) circle[];
			\fill[extrasample] (7.5, 1.9) coordinate (s16) circle[];

		\end{tikzpicture}
		\caption{Safety}
		\label{example:unsafe}
	\end{subfigure}
	\begin{subfigure}[c]{\imagewidth}
		\begin{tikzpicture}[shorten >=1pt, scale=.65, yscale=.6, xscale=0.6, auto]

			\draw[->] (0, 0) --++ (0, 5.0) node[anchor=north east]{$x$};
			\draw[->] (0, 0) --++ (8.0, 0) node[anchor=north]{$t$};
			\draw[dashed, color=red, semithick] (0, 4) --++ (8.0, 0);

			\newcommand{\uncertainsample}[3]{\fill[uncertainsample] (#1-.1, #2-#3) rectangle (#1+.1, #2+#3);}
			
			\uncertainsample{0}{2}{0.5};
			\uncertainsample{2}{2.1}{0.4};
			\uncertainsample{3}{1.5}{0.6};
			\uncertainsample{4}{0.9}{0.3};
			\uncertainsample{5}{2.2}{0.5};
			\uncertainsample{7}{1.9}{0.4};
		\end{tikzpicture}
		\caption{Uncertain}
		\label{example:uncertain}
	\end{subfigure}
	\caption{Monitoring at discrete time steps~\cite{GA22}}
\end{figure*}

\paragraph{Monitoring using uncertain linear systems as a bounding model}
A challenge when performing offline monitoring is to ``recreate'' or guess the samples at the missing time steps.
That is, when the system under monitoring is a black-box, with a log in the form of an aperiodic timed sequence of valuations of continuous variables (with missing valuations at various time steps): \textit{how to be certain that in between two discrete valuations the specification was not violated at another discrete time step at which no logging was performed?}
Consider \cref{example:full}, a system for which a logging occurs at every discrete time step.
When logging occurs at only \emph{some} time steps (due to some sensor faults, or to save energy with only a sparse, scattered logging), a possible output is in \cref{example:monitored}.
In such a setting,
\emph{how to be certain that, in between two discrete samples, another discrete sample (not recorded) did not violate the specification?}
That is, in \cref{example:monitored}, there is no way to formally guarantee that the unsafe zone (\ie{} above the red, dashed line) was never reached by another discrete sample which was not recorded.
In many practical cases, a piecewise-constant or linear approximation (see, \eg{} \cref{example:linear,example:constant}, where the large blue dots denote actual samples, while the small green dots denote reconstructed samples using some extrapolation) is arbitrary and not appropriate---it can yield a ``safe'' answer, while the actual system could have actually been unsafe at some of the missing time steps.
On the contrary, assuming a completely arbitrary dynamics will always yield ``potentially unsafe''---thus removing the interest of monitoring.
Without any knowledge of the model, one can always assume that the behavior given in \cref{example:unsafe} could happen.
This behavior shows that the variable~$x$ is suddenly crossing the unsafe region (dashed) at some unlogged discrete time step---even though this is unlikely if the dynamics is known to vary ``not very fast''.
To alleviate such issues, we proposed in~\cite{GA22} an offline monitoring algorithm using a \emph{bounding model}, \ie{} a rough overapproximation of the system behavior (originally introduced in \cite{WAH22TCPS} in a different context). The proposed method is based on the reachable set computation of uncertain linear systems~\cite{ghosh1} that can detect safety violations with limited false alarms. %

We also considered in~\cite{GA22} an \emph{online} monitoring algorithm, aiming at energetic efficiency, by recording samples only when required (\ie{} when the system may get closer to a violation).
\ourTool{} implements both our offline and online monitoring algorithms~\cite{GA22}.
We also provide here the steps to easily recreate the results of the two case studies in~\cite{GA22}. %

\paragraph{Experimental setting}
Given an aperiodic (\ie{} missing valuations at various time steps) and a noisy log (\ie{} the valuations that are present in the log can have an added noise---an overapproximation of the actual state), \ourTool{} can perform offline monitoring of the system to detect safety of the system behavior.
Further, \ourTool{} can be used in an online setting to log only when necessary---thus targeting energy efficiency while logging. \ourTool{} can be run on a standard laptop with a Linux operating system (see the installation guide for details~\cite{install_guide}). The details of how to use the tool, with illustrative examples, can be found in the user guide~\cite{user_guide}.

\paragraph{Outline} %
\cref{section:software-description} describes the software;
\cref{section:architecture} describes the architecture;
\cref{section:functionalities} describes the functionalities;
\cref{section:illustrative} describes two illustrative examples from medical and automotive domains, with necessary steps to recreate their results;
\cref{section:impact} discusses the impact of \ourTool{};
\cref{section:related} briefly reviews related works;
\cref{section:conclusion} concludes and discusses future research.
\section{Software Description}\label{section:software-description}
\ourTool{} is an open-source software, implemented in Python, running on Linux platforms.
Our experiments~\cite{GA22} suggest that \ourTool{} is able to perform monitoring of reasonably large systems in a reasonable time.
For example, we are able to monitor a five-dimensional system (\ie{} a system with 5 continuous variables monitored at the same time) for 2000 time steps, with only actual 300 samples (note that, the fewer samples, the higher is the monitor computation time---as it is required to ``recreate'' the missing samples using reachable sets) in under 2.5 minutes on a standard laptop.
We believe that \ourTool{} will not just be helpful to engineers to analyze logs to detect safety violations in several areas of research (such as in robotics to detect collision and other undesirable behaviors), but also to researchers to further develop monitoring-based approaches---in that case \ourTool{} can be used for comparison.

\section{Software Architecture}\label{section:architecture}

\begin{figure}[tb!]
  \centering
\includegraphics[width=\textwidth]{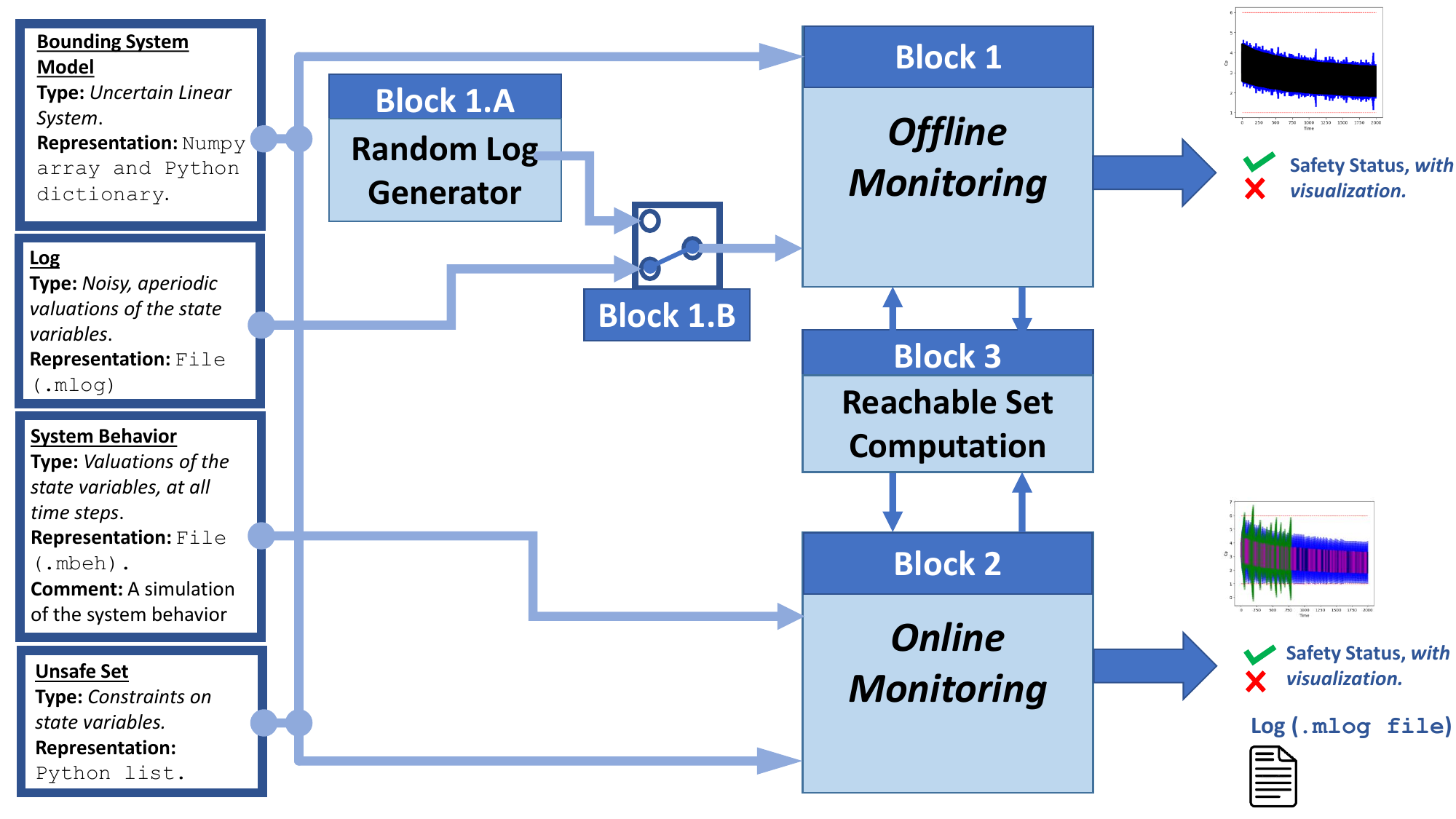}
 \caption{\ourTool{} \textbf{\textit{Architecture.}} Each block identifies a core part of the tool, and the arrows indicate the flow of data.}
 \label{fig:architecture}
\end{figure}

\begin{figure}[tb!]
  \centering
\includegraphics[width=\textwidth]{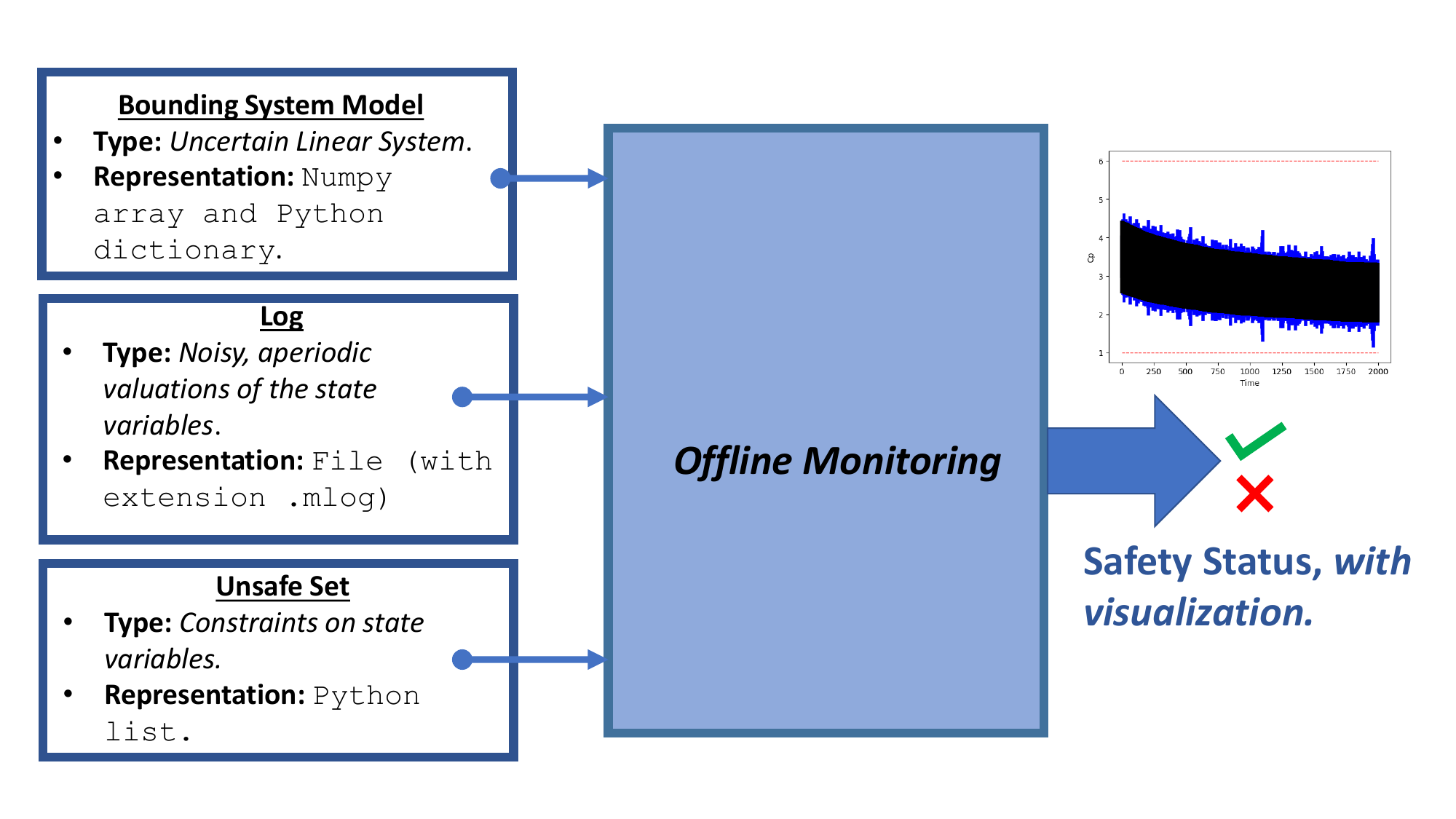}
 \caption{Dataflow diagram of the offline monitoring functionality of \ourTool{}.}
 \label{fig:io_offline}
\end{figure}

\begin{figure}[tb!]
  \centering
\includegraphics[width=\textwidth]{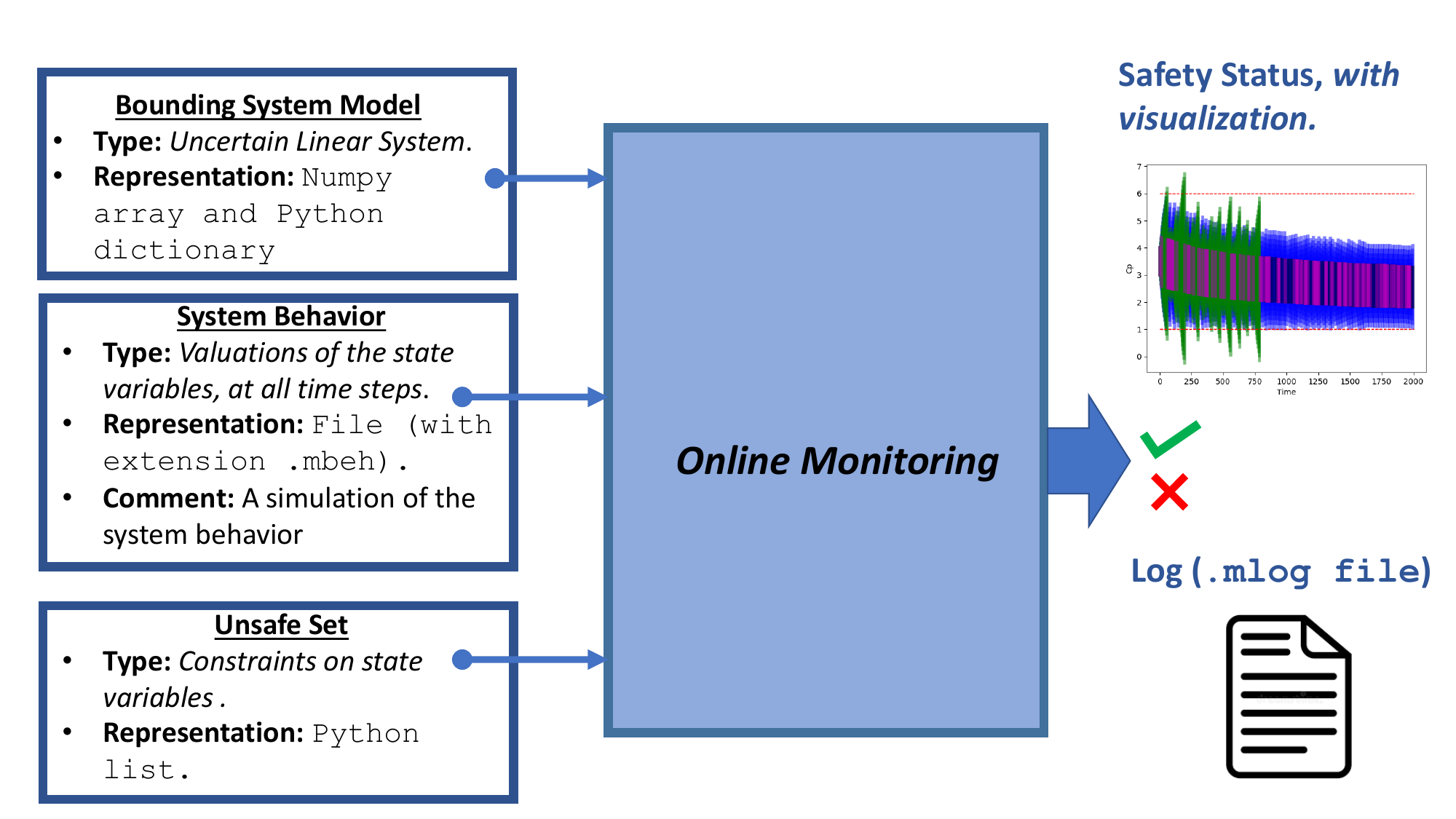}
 \caption{Dataflow diagram of the online monitoring functionality of \ourTool{}.}
 \label{fig:io_online}
\end{figure}

The architecture of \ourTool{} is given in \cref{fig:architecture}. Each block in \cref{fig:architecture} represents a core part (either functional or input) of the tool, and the arrows indicate the flow of data. 
The dataflow of \ourTool{}, for both \cref{fig:io_offline,fig:io_online} (offline and online monitoring respectively), starts from the blocks in the left (\eg{} bounding model, unsafe set, etc.)\ and ends at the extreme right of the figures (outputting the safety status, visualization, and/or synthesized log).
Since both the offline (\texttt{Block~1} of \cref{fig:architecture,fig:io_offline}) and online monitoring algorithm (\texttt{Block~2}  in \cref{fig:architecture,fig:io_online}) uses reachability of uncertain linear systems (\texttt{Block~3} of \cref{fig:architecture}), a data exchange occurs between \texttt{Block~1} and \texttt{Block~3}, as well as \texttt{Block~2} and \texttt{Block~3}.
\ourTool{} implements a built-in reachability algorithm.
The native reachable set computation support facilitates faster computing (\ie{} no data exchange with third party tool is required, which would potentially require additional data reformatting) with no additional tool installation required.
\ourTool{} employs an algorithm proposed in~\cite{ghosh1} to compute the reachable set of uncertain linear systems. The algorithm first computes the reachable set of the nominal dynamics (which excludes uncertainties) and then computes the reachable set related to the uncertainties in the dynamics. These two sets are then combined using the Minkowski sum to obtain the reachable set of the entire dynamics. Although computing the reachable set of the nominal dynamics is straightforward, the reachable set related to uncertainties is challenging to compute.
After obtaining the reachable sets, \ourTool{} verifies the safety of these sets by comparing them against provided safety specifications. These safety specifications are constraints on state variables, which can be complex and involve multiple state variables (\eg{} linear inequalities involving several state variables). To represent such safety specifications, \ourTool{} uses a special type of polytope called \emph{zonotopes}, which can be expressed as an affine transformation of a unit box.
\ourTool{} can check multiple safety specifications, each represented as a zonotope and involving several state variables.

\subsection{Implementation}
\label{subsec:implementation}

\ourTool{} is implemented using \texttt{Python 3.7.x}, and runs in a Linux environment.
The architecture is given in \cref{fig:architecture}.

The tool can be used in the following two ways.
On the one hand, users can use it through the provided virtual machine\footnote{\href{https://www.doi.org/10.5281/zenodo.7888502}{\nolinkurl{10.5281/zenodo.7888502}}}, which already contains all the necessary dependencies and has the path variable set. Nevertheless, users are still required to obtain and install the Gurobi license themselves, since Gurobi only grants free academic licenses to individuals.
On the other hand, the tool can also be downloaded and setup from its public GitHub repository\footnote{{\scriptsize\url{https://github.com/bineet-coderep/MoULDyS/releases/tag/v1.1}}}. If users aim to recreate the results in the paper or simply employ it for basic monitoring purposes, using the provided virtual machine is recommended. However, if the tool will be used for research and development purposes, it is recommended to download and set it up on a local machine. The detailed installation instructions are provided in~\cite{install_guide}.
The system model (represented as an uncertain linear system), in \ourTool{}, is represented with a \texttt{numpy} array and a dictionary. The unsafe set is represented with a Python list.
An example code of how to encode the system model and its unsafe specification can be found in the public repository\footnote{\url{https://www.github.com/bineet-coderep/MoULDyS/blob/main/src/tutorial/TutorialOfflineMonitoring.py}}.
The log and the system behavior, on the other hand, is given as a file to \ourTool{} (\ourTool{} can also generate random logs; discussed later). Logs can either be represented as zonotopes or intervals (an example of the required file can be found in \ourTool{} public repository\footnote{\url{https://www.github.com/bineet-coderep/MoULDyS/blob/main/data/toyEg_5_interval.mlog}}).

Both the online and offline monitoring algorithm (\texttt{Block~1} and \texttt{Block~2} of \cref{fig:architecture}) have been implemented in Python using standard libraries, namely \texttt{numpy}, \texttt{scipy} and \texttt{mpmath}. 
The reachable set computation (\texttt{Block~3} of \cref{fig:architecture}) module implements the algorithm proposed in~\cite{ghosh1} in Python.
Both the online and the offline module require performing intersection checking of zonotopes~\cite{GA22}, which has been implemented as an optimization formulation using \texttt{Gurobi}. \texttt{Gurobi} has been further used to visualize the reachable sets.

\paragraph{Installation}
While the virtual machine comes preinstalled with the required dependencies, setting up \ourTool{} on a local machine requires installing the following dependencies: \texttt{numpy}, \texttt{scipy}, \texttt{mpmath}, \texttt{pandas}, \texttt{Gurobi} along with \texttt{gurobipy} (\texttt{Gurobi} requires an \emph{ad-hoc} install but is free to use for academic purposes).
The detailed steps for installing \ourTool{} are given in the installation guide~\cite{install_guide}.

\paragraph{User Guide} A tutorial on how to use several functionalities of \ourTool{}, along with sample codes (encoding a toy dynamics), is given in the user guide~\cite{user_guide}.

\section{Software Functionalities}\label{section:functionalities}
In this section, we discuss the core functionalities of \ourTool{}:
\begin{description}
    \item [Offline Monitoring] The offline monitoring requires the bounding model of the system (represented as an uncertain linear system) to be given as input. Further, a log of the system behavior is required, which can be achieved by the following ways:
    \begin{ienumerate}
        \item If the user already has a log to be monitored, it can be simply passed as an input to \texttt{Block~1} of \cref{fig:architecture}.
        \item Alternatively, \ourTool{} can also generate a random (noisy and aperiodic) log of the system, from a given initial set, using \texttt{Block~1.A}. The selection between the two possible choices is facilitated by \texttt{Block~1.B}.
        Analyzing the log, the final output of the offline monitoring is either \texttt{safe} (indicating the system behavior is certainly safe at all time steps), or \texttt{possibly-unsafe} (indicating the system might have shown unsafe behavior). An example code snippet to perform offline monitoring, on a toy example, can be found in its public repository.\footnote{\url{https://www.github.com/bineet-coderep/MoULDyS/blob/main/src/tutorial/TutorialOfflineMonitoring.py}}
    \end{ienumerate}
    \item[Online Monitoring] The online monitoring requires the bounding model of the system (represented as an uncertain linear system) to be given as input, as well as the actual behavior of the system.
    The actual behavior of the system is given as a file (with extension \texttt{.mbeh}) representing the values of the state variables at every time step---an example of the expected file (with extension \texttt{.mbeh}), containing valuation of the state variables at every time step, can be found in its public repository.\footnote{\url{https://www.github.com/bineet-coderep/MoULDyS/blob/main/data/toyEg_5_interval.mbeh}}
    The final output of this feature is the safety status of the system (\texttt{safe}/\texttt{possibly-unsafe}), and a synthesized log. An example code snippet to perform online monitoring, on a toy example, can be found in its public repository.\footnote{\url{https://www.github.com/bineet-coderep/MoULDyS/blob/main/src/tutorial/TutorialOnlineMonitoring.py}}
\end{description}

As a side functionality, \ourTool{} also allows to generate random logs (using \texttt{Block~1.A}).
While this is not strictly speaking part of monitoring, it greatly helps to perform experiments using \ourTool{}.
Basically, given a bounding model, \ourTool{} can generate a random log following the bounding model.
\section{Illustrative examples}\label{section:illustrative}

In this section, we briefly recall the two case studies presented in \cite{GA22} that use a prototype version of \ourTool{}.
The two case studies, automated anesthesia delivery and adaptive cruise control, demonstrate the applicability and usability of \ourTool{}. Further, we provide detailed steps to recreate the results presented in \cite{GA22} using \ourTool{}. 
\begin{description}
    \item[Anesthesia] \cite{GDM14} presents an automated anaesthesia delivery model, with the drug propofol. The system models the metabolization of the drug by the body, and the depth of hypnosis. The state variables encode the various concentration levels---that must be within a certain limit at all times---modeling the metabolization of the drug and the depth of hypnosis.
    Note that a higher concentration level would mean that the patient remains unconscious for a longer period of time, while a lower concentration level would mean that patient remains conscious during the surgery---which can be traumatic.
    \ourTool{} can help performing an automated monitoring of the patients without compromising on safety. %

    \item[Adaptive Cruise Control (ACC)]
	\cite{NHBCAGOPT16} presents a model of ACC with state variables as velocity, distance between two vehicles, and the velocity of the lead vehicle.
	Offline monitoring provides an automated way to detect the \emph{cause of the crash} and \emph{who was at fault}.
	Similarly, consider a vehicle driving on a highway with a vehicle in its sight. The ACC unit will have to continuously read sensor values to track several parameters, such as acceleration of the lead vehicle, braking force, etc.---causing a waste of energy.
	In these cases, deploying online monitoring on the vehicle ACC will ensure that the sensor values are only read when there is a potential unsafe behavior---thus saving energy.
	\cite{GA22}~provides several such practical cases where monitoring is useful---in \cite[Figs.~5 and~6]{GA22} (also recalled in the appendix in \cref{fig:offlineACC,fig:onlineACC} respectively).
\end{description}

The case studies mentioned above study the following aspects with regards to monitoring:
\begin{ienumerate}
    \item Impact of number of samples in the log.
    \item Impact of uncertainties in the samples of the log.
    \item Demonstrating online monitoring.
    \item Comparing offline and online monitoring.
\end{ienumerate}
In the following, we provide the detailed steps to recreate the results in \cite[Section 5]{GA22}. In particular, the results that we wish to recreate, from~\cite{GA22}, are given in \cref{fig:offlinePKPD,fig:onlinePKPD,fig:offlineACC,fig:onlineACC}.

\subsection*{Recreating Results}
\label{sec:recreate}
The results of the Anesthesia case study and the ACC case study can be recreated by using the scripts provided in the GitHub repository\footnote{\scriptsize\url{https://www.github.com/bineet-coderep/MoULDyS/tree/main/src/recreate_results_from_paper}}.
The detailed steps to recreate the results from both the case studies are given in~\cite{rec_guide}.

Note that, in~\cite[Section 5]{GA22}, logs were randomly generated \emph{during our experiments}, and therefore the results from \cite[Section 5]{GA22} cannot be \emph{stricto sensu} be recreated, as their reproducibility script re-generates a random log, which may differ from the one actually shown in \cite[Section 5]{GA22}.

Therefore, we modified our scripts so that the log is given as an input, on which monitoring is performed:
we thus generated the logs statically, and embedded them in the reproducibility capsule, in order for users to reproduce exactly the result we present here.
These logs were generated with the same logging probabilities (and initial sets) as in~\cite[Section 5]{GA22}.
In the rest of the section, we discuss the steps for replicating the new \cref{fig:offlinePKPD,fig:onlinePKPD,fig:offlineACC,fig:onlineACC}.

\paragraph{Anesthesia} The main results of the Anesthesia case study are provided in \cref{fig:offlinePKPD,fig:onlinePKPD}
(variant of~\cite[Figures~3 and~4]{GA22}).
We use \texttt{python Anesthesia.py -offline i}, with $\mathtt{i} \in \{1,2,3,4\}$ to recreate \cref{fig:anesthesia_fig1,fig:anesthesia_fig2,fig:anesthesia_fig3,fig:anesthesia_fig4} respectively.
To recreate \cref{fig:anesthesia_online}, we use \texttt{python~Anesthesia.py~-online}.
To recreate \cref{fig:anesthesia_comp}, we use \texttt{python~Anesthesia.py~-compare}.
The \texttt{Anesthesia.py} script is provided in its GitHub repository.\footnote{\scriptsize\url{https://www.github.com/bineet-coderep/MoULDyS/blob/main/src/recreate_results_from_paper/Anesthesia.py}}

\paragraph{ACC} The main results of the ACC case study are provided in \cref{fig:offlineACC,fig:onlineACC} (variant of~\cite[Figures~5 and~6]{GA22}).
The results of the ACC case study can be recreated in a similar manner to the Anesthesia case study, by simply using the \texttt{ACC.py}\footnote{\scriptsize\url{https://github.com/bineet-coderep/MoULDyS/blob/main/src/recreate_results_from_paper/ACC.py}} script instead of the \texttt{Anesthesia.py} script.

\section{Impact}\label{section:impact}

While \ourTool{} remains a prototype based on a recent algorithm~\cite{GA22}, and is by no means a widely used software in an industrial context for the time being, we believe it has an interesting potential to gain up a user base interested in monitoring black-box cyber-physical systems against safety properties.
To the best of our knowledge, it is the first software allowing monitoring logs featuring not only uncertainty (due to sensor's imperfect behavior) but also potentially missing samples, together with a bounding model going beyond the class of linear systems.
This is in contrast with, \eg{} \cite{WAH22TCPS} in which our bounding model was restricted to linear models.

In addition, \ourTool{} can perform \emph{online} monitoring, with a focus on energetic efficiency: by triggering a sample recording only when necessary (\ie{} when \ourTool{} informs the system that it may get close to an unsafe behavior according to its online algorithm), the system saves energy, \ie{} only records samples (which needs network bandwidth usage, as well as processor and memory usage) when necessary instead of at every time unit.

Our applications recalled in \cref{section:illustrative} show that \ourTool{} can be applied to challenging domains such as health and autonomous driving, giving interesting results (\eg{} limited number of false alarms) in a reasonable execution time making it suitable to real-time applications.

\section{Related works}\label{section:related}

Monitoring complex systems, and notably cyber-physical systems, drew a lot of attention in the last decades~\cite{BDDFMNS18}.
We briefly review close works in the following.

\textsc{MonPoly}~\cite{BKZ17} is a monitoring tool taking as specification formulas expressed using MFOTL (metric first-order temporal logic).
It is entirely black-box: the only input beyond the formula is the \emph{log}, \ie{} a sequence of timestamped system events, potentially with numeric arguments (\eg{} ``\texttt{@10 withdraw (Alice,6000)}'', expressing that a withdrawal occurs at timestamp~10).

In~\cite{MCW21}, the focus is on online monitoring over real-valued signals, using MTL as the specification formalism.
Again, the system is black-box.

In~\cite{WAH23ToSEM}, parametric timed pattern matching is made, on an entirely black-box system, \ie{} without any prior knowledge of the system; the tools used are \imitator{}~\cite{Andre21} and a prototypal tool \textsf{ParamMONAA}.
The output is a set of intervals where a property is valid/violated, possibly with a set of timing parameter valuations.

In~\cite{WAH22TCPS}, we proposed \emph{model-bounded monitoring}: instead of monitoring a black-box system against a sole specification, we use in addition a (limited, over-approximated) knowledge of the system, to eliminate false positives.
This over-approximated knowledge is given in~\cite{WAH22TCPS} in the form of a \emph{linear hybrid automaton} (LHA)~\cite{HPR94}.
We use in~\cite{WAH22TCPS} both an \emph{ad-hoc} implementation, and another one based on PHAVerLite~\cite{BZ19}.
In this work, we share with~\cite{WAH22TCPS} the principle of using an over-approximation of the model to rule out some violation of the specification, which comes in contrast with the aforementioned works.
However, we consider here a different formalism, and we work on discrete samples.
In terms of expressiveness of the over-approximated model, while our approach can be seen as less expressive than~\cite{WAH22TCPS}, in the sense that we have a single (uncertain) dynamics (as opposed to LHAs, where a different dynamics can be defined in each mode), our dynamics is also significantly \emph{more expressive} than the LHA dynamics of~\cite{WAH22TCPS}; we consider not only the class of linear dynamical systems, but even fit into a special case of non-linear systems, by allowing \emph{uncertainty} in the model dynamics.

In~\cite{MP16,MP18}, a monitor is constructed from a system model in differential dynamic logic~\cite{Platzer12}.
The main difference between~\cite{MP16,MP18} and our approach relies in the system model: in~\cite{MP16,MP18}, the compliance between the model and the behavior is checked at runtime, while our model is assumed to be an over-approximation of the behavior---which is by assumption compliant with the model.

\section{Conclusion and Future Work}\label{section:conclusion}

Monitoring black-box complex cyber-physical systems can be delicate, and may lead to false alarms.
\ourTool{} is a Python-based tool implementing offline and online monitoring algorithms.
A first crux of \ourTool{} is to be able to manage logs with uncertainty over the logged state variables, as well as missing samples.
A second crux is the use of a bounding model in the form of uncertain linear systems, helping to reduce the number of false alarms.
\ourTool{} can analyze logs efficiently to detect possible safety violations that might have caused an unsafe behavior.
Further, \ourTool{} can also be used in an online setting where the system is sampled only when there is a risk of safety violation.
As a result, the online monitoring is able to decrease the number of samples, therefore reducing energy consumption at runtime.
\ourTool{} is available under the GNU General Public License.

In future, we wish to extend \ourTool{} to support %
uncertainty not only in the log valuations (the value of a sensor at a given timestamp), but also uncertainty in the log timestamps themselves: this makes sense when some sensors are distributed with drifting clocks, or when network delays make the exact recording timestamp imprecise.

\section*{Acknowledgments}
Bineet Ghosh was supported by the National Science Foundation (NSF) of the United States of America under grant number 2038960.
This work is partially supported by the ANR-NRF French-Singaporean research program ProMiS (ANR-19-CE25-0015 / 2019 ANR NRF 0092)
and by ANR BisoUS (ANR-22-CE48-0012).

\newcommand{\CCIS}{Communications in Computer and Information Science}
\newcommand{\ENTCS}{Electronic Notes in Theoretical Computer Science}
\newcommand{\FAC}{Formal Aspects of Computing}
\newcommand{\FundInf}{Fundamenta Informaticae}
\newcommand{\FMSD}{Formal Methods in System Design}
\newcommand{\IJFCS}{International Journal of Foundations of Computer Science}
\newcommand{\IJSSE}{International Journal of Secure Software Engineering}
\newcommand{\IPL}{Information Processing Letters}
\newcommand{\JAIR}{Journal of Artificial Intelligence Research}
\newcommand{\JLAP}{Journal of Logic and Algebraic Programming}
\newcommand{\JLAMP}{Journal of Logical and Algebraic Methods in Programming} %
\newcommand{\JLC}{Journal of Logic and Computation}
\newcommand{\LMCS}{Logical Methods in Computer Science}
\newcommand{\LNCS}{Lecture Notes in Computer Science}
\newcommand{\RESS}{Reliability Engineering \& System Safety}
\newcommand{\STTT}{International Journal on Software Tools for Technology Transfer}
\newcommand{\TCS}{Theoretical Computer Science}
\newcommand{\ToPNoC}{Transactions on Petri Nets and Other Models of Concurrency}
\newcommand{\TSE}{{IEEE} Transactions on Software Engineering}

\renewcommand*{\bibfont}{\small}
\printbibliography[title={References}]

\appendix

\section{\texttt{MoULDyS}: A Monitoring Tool for Autonomous Systems}
\label{sec:mouldys}
\begin{figure}[tb!]
  \centering
\includegraphics[width=\textwidth]{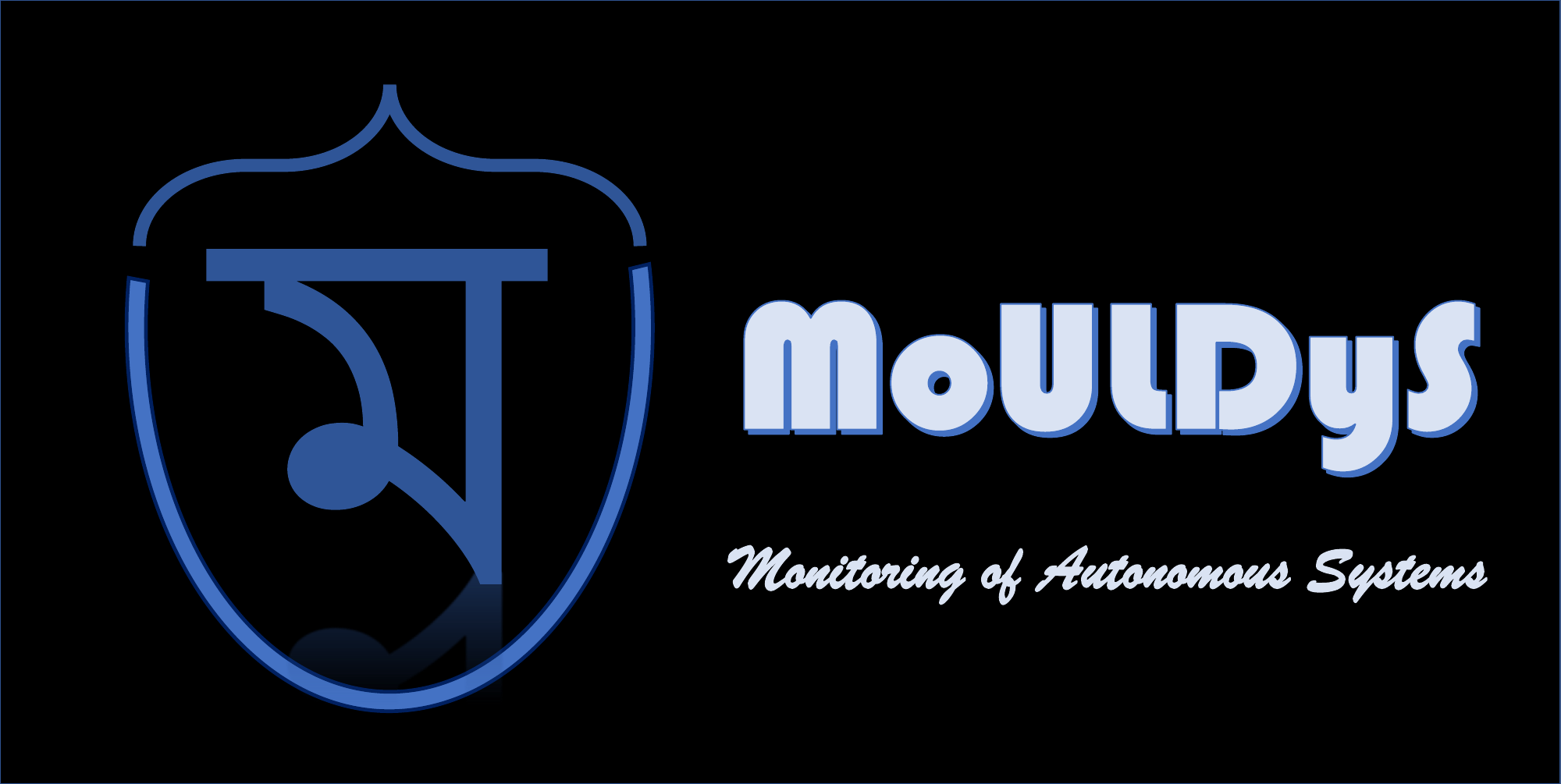}
 \caption{\ourTool{} Logo.}
 \label{fig:logo}
\end{figure}

The various details of \ourTool{} are given as follows:
\begin{description}
    \item[Logo] The tool logo is given in \cref{fig:logo}.
    \item[Webpage] The tool webpage can be found here: \url{https://www.sites.google.com/view/mouldys}.
    \item[Code] \ourTool{} is an open-source tool under the \license{} license. The code can be found in a public GitHub repository: {\scriptsize\url{https://github.com/bineet-coderep/MoULDyS/releases/tag/v1.1}}.
    \item[Installation Guide] The installation guide is available in~\cite{install_guide}.
    \item[User Guide] The user guide is available in~\cite{user_guide}.
    \item[Result Recreation Guide Guide] A prototype version of \ourTool{} was used to perform the experiments in~\cite{GA22}.
		The detailed steps to recreate the results in~\cite{GA22}, through easy-to-use scripts, are available in~\cite{rec_guide}.
    
\end{description}

\section{Recreating Experimental Results}
\label{sec:rec_figs}
\begin{figure*}
    \begin{subfigure}{.45\textwidth}
        \centering
        \includegraphics[width=\linewidth]{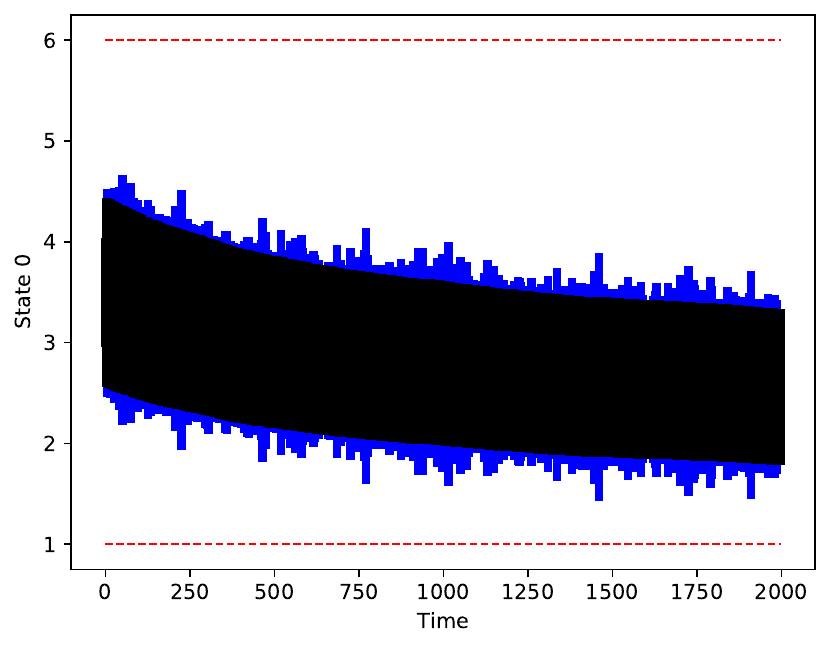}
        \caption{Monitoring with frequent samples, and low uncertainty}
        \label{fig:anesthesia_fig3}
    \end{subfigure}
    \hfill
    \begin{subfigure}{.45\textwidth}
        \centering
        \includegraphics[width=\linewidth]{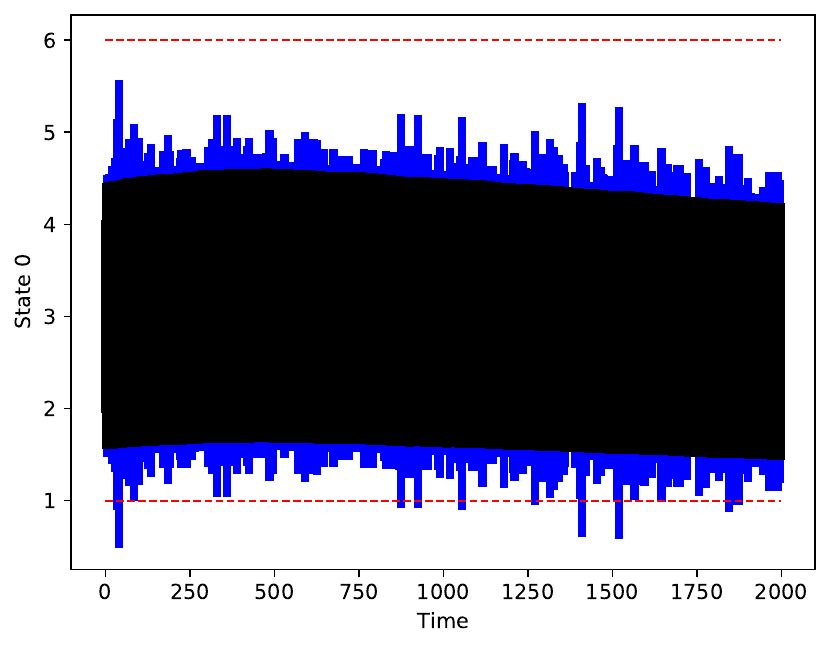}
        \caption{Monitoring with frequent samples, and high uncertainty}
        \label{fig:anesthesia_fig4}
    \end{subfigure}
    
    \vfill
    
    \begin{subfigure}{.45\textwidth}
        \centering
        \includegraphics[width=\linewidth]{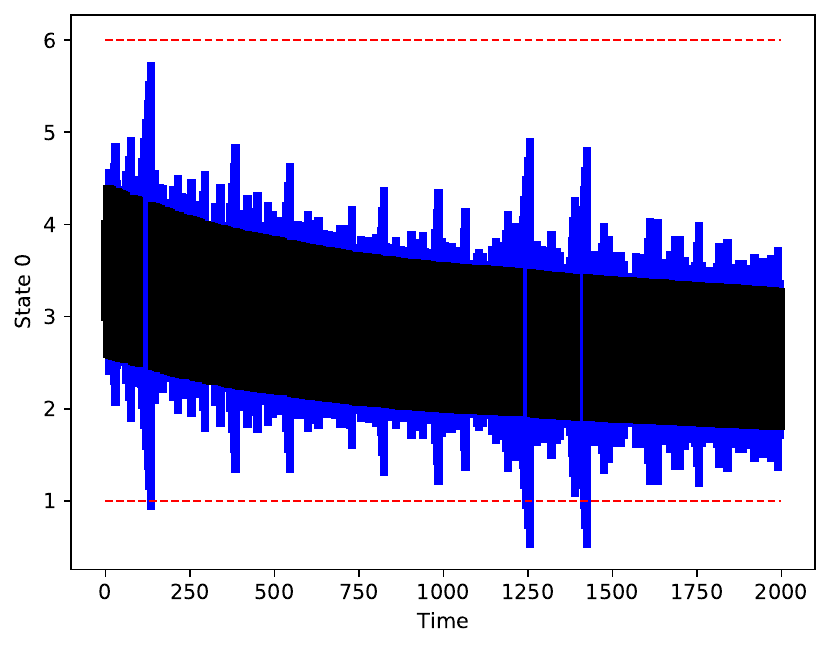}
        \caption{Monitoring with sporadic samples, and low uncertainty}
        \label{fig:anesthesia_fig1}
    \end{subfigure}
    \hfill
    \begin{subfigure}{.45\textwidth}
        \centering
        \includegraphics[width=\linewidth]{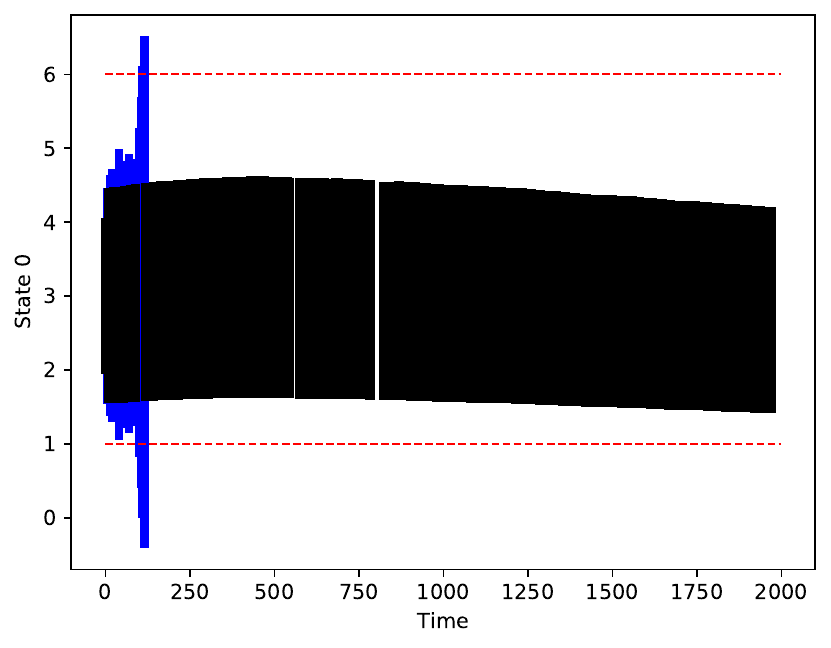}
        \caption{Monitoring with sporadic samples, and high uncertainty}
        \label{fig:anesthesia_fig2}
    \end{subfigure}
    \caption{\textbf{\textit{Offline Monitoring (Anesthesia).}} We plot the change in concentration level of $c_p$ with time. The volume of the samples increases from left to right, and the probability of logging increases from bottom to top. The blue regions are the reachable sets showing the over-approximate reachable sets as computed by the offline monitoring, the black regions are the samples from the log given to the offline monitoring algorithm, and the red dotted line represents safe distance level. Note that although \textbf{Figure~1} and \textbf{Figure~4} (\cref{fig:anesthesia_fig1,fig:anesthesia_fig4}) reachable sets' seem to intersect with the red line (unsafe set), the refinement module infers them to be \emph{unreachable}, therefore concluding the system behavior as \emph{safe}---unlike \cref{fig:anesthesia_fig2}. These plots are a stochastic recreation of~\cite[Figure 3]{GA22}.} 
    \label{fig:offlinePKPD}
\end{figure*}

\begin{figure*}
    \begin{subfigure}{.45\textwidth}
        \centering
        \includegraphics[width=\linewidth]{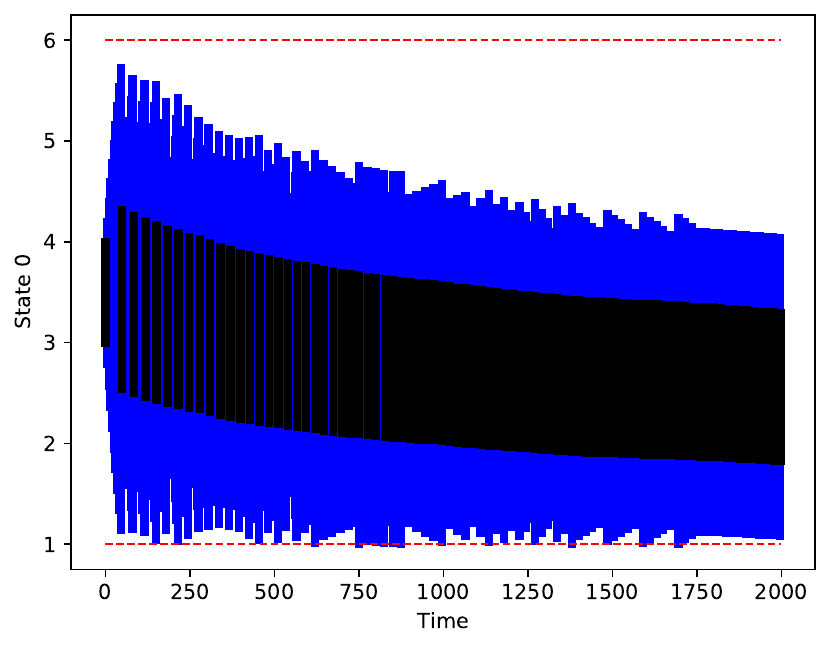}
        \caption{\textbf{Online Monitoring}}
        \label{fig:anesthesia_online}
    \end{subfigure}
    \hfill
    \begin{subfigure}{.45\textwidth}
        \centering
        \includegraphics[width=\linewidth]{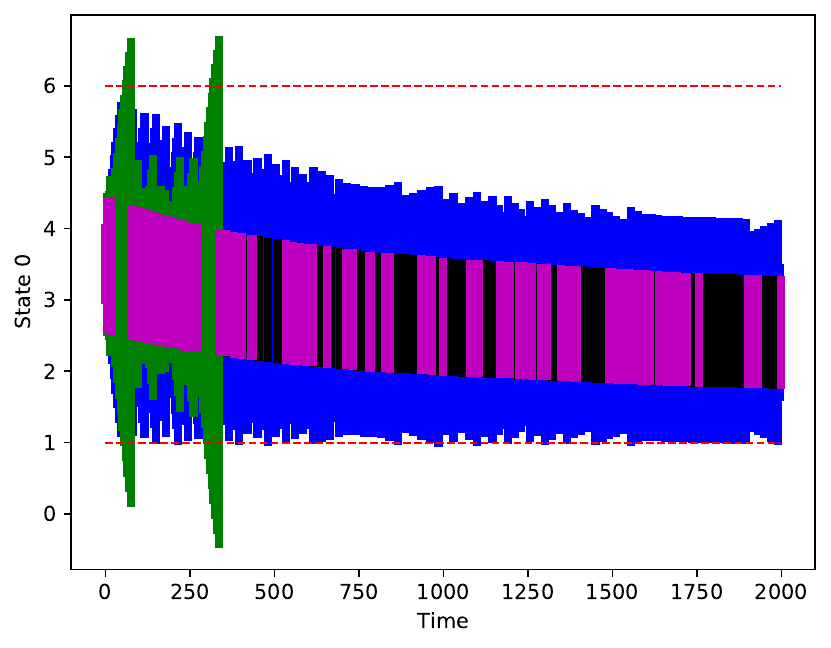}
        \caption{\textbf{Compare Online and Offline Monitoring}}
        \label{fig:anesthesia_comp}
    \end{subfigure}
    \caption{\textbf{\textit{Online Monitoring (Anesthesia).}} We plot the change in concentration level of~$c_p$ with time. The blue regions are the reachable sets showing the over-approximate reachable sets as computed by the online monitoring, the black regions are the samples generated when the logging system was triggered by the online monitoring algorithm, and the red dotted line represents safe concentration levels. \textit{Online Monitoring (\cref{fig:anesthesia_online}):} We apply our online monitoring to the anesthesia model. \textit{Compare (\cref{fig:anesthesia_comp}):} We compare our online and offline algorithms. The green regions are the reachable sets showing the over-approximate reachable sets between two consecutive samples from the offline logs, the magenta regions are the offline logs, given as an input to the offline monitoring algorithm, generated by the logging system, and the red dotted line represents safe concentration levels. The blue regions are the reachable sets showing the over-approximate reachable sets as computed by the online monitoring, the black regions are the samples generated when the logging system was triggered by the online monitoring algorithm, and the red dotted line represents safe concentration levels. These plots are a stochastic recreation of~\cite[Figure 4]{GA22}.} 
    \label{fig:onlinePKPD}
\end{figure*}

\begin{figure*}
    \begin{subfigure}{.45\textwidth}
        \centering
        \includegraphics[width=\linewidth]{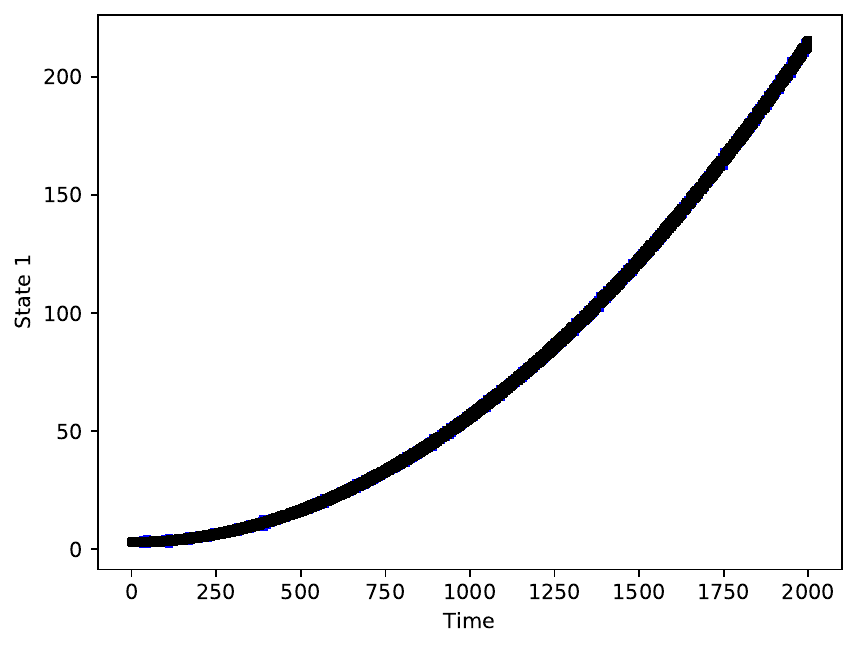}
        \caption{Monitoring with frequent samples, and low uncertainty}
        \label{fig:acc_fig3}
    \end{subfigure}
    \hfill
    \begin{subfigure}{.45\textwidth}
        \centering
        \includegraphics[width=\linewidth]{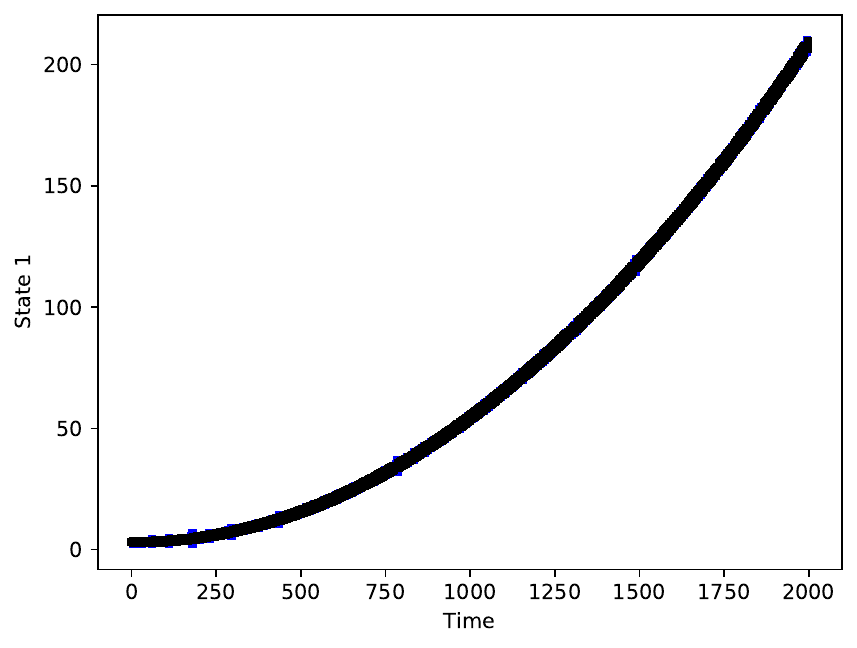}
        \caption{Monitoring with frequent samples, and high uncertainty}
        \label{fig:acc_fig4}
    \end{subfigure}
    
    \vfill
    
    \begin{subfigure}{.45\textwidth}
        \centering
        \includegraphics[width=\linewidth]{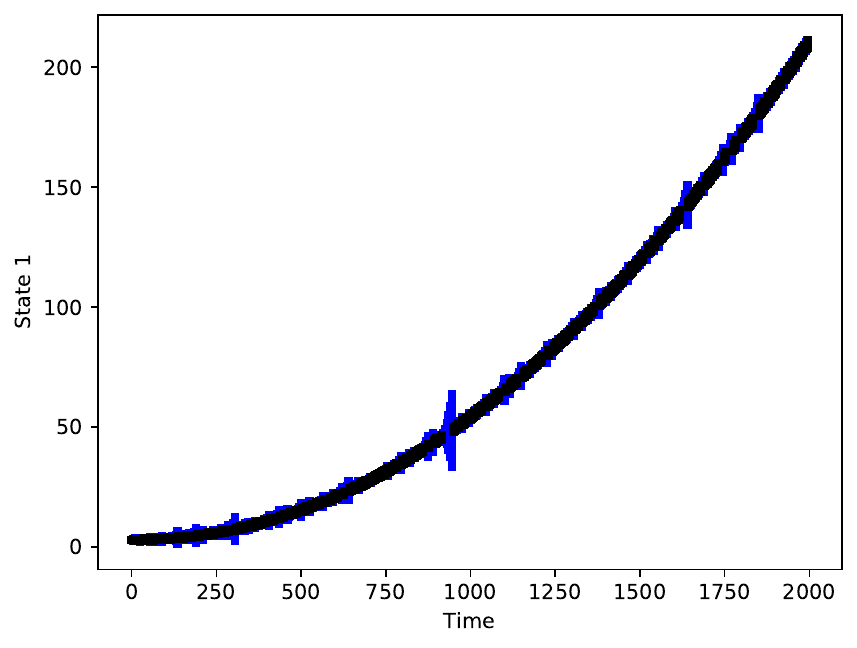}
        \caption{Monitoring with sporadic samples, and low uncertainty}
        \label{fig:acc_fig1}
    \end{subfigure}
    \hfill
    \begin{subfigure}{.45\textwidth}
        \centering
        \includegraphics[width=\linewidth]{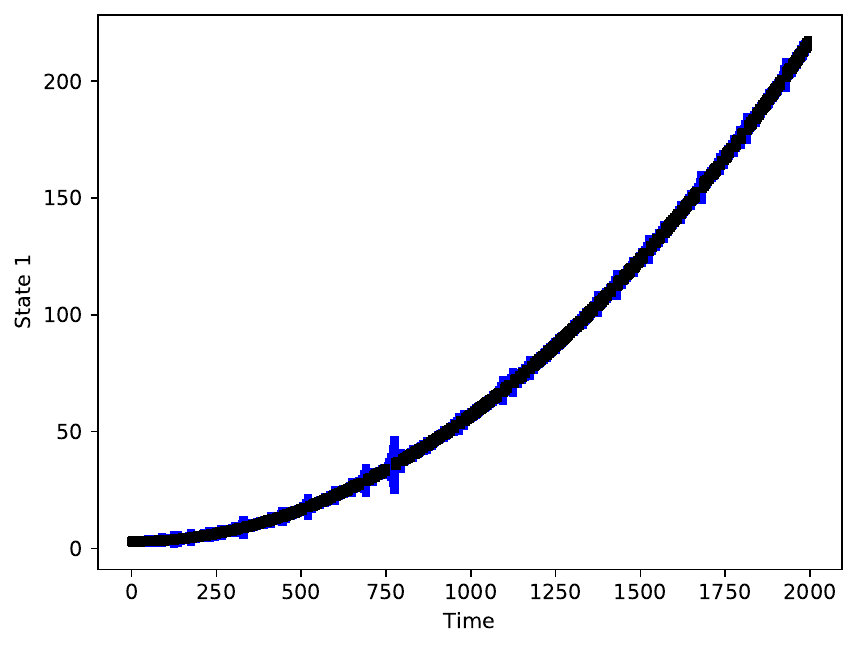}
        \caption{Monitoring with sporadic samples, and high uncertainty}
        \label{fig:acc_fig2}
    \end{subfigure}
    \caption{\textbf{\textit{Offline Monitoring (ACC).}} We plot the change in distance $h$ between the vehicles with time. The volume of the samples increases from left to right, and the probability of logging increases from bottom to top. These plots are a stochastic recreation of~\cite[Figure 5]{GA22}} 
    \label{fig:offlineACC}
\end{figure*}

\begin{figure*}
    \begin{subfigure}{.45\textwidth}
        \centering
        \includegraphics[width=\linewidth]{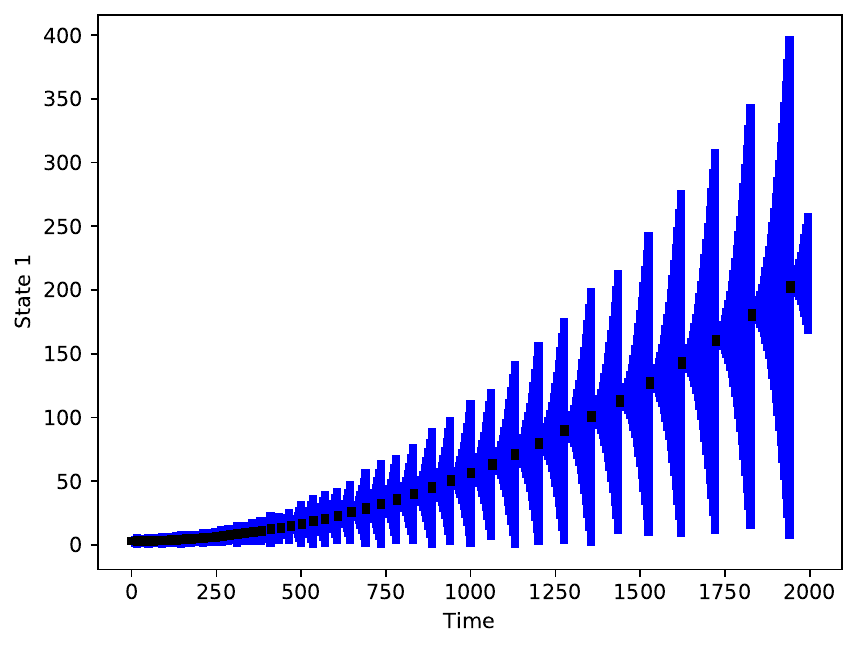}
        \caption{\textbf{Online Monitoring}}
        \label{fig:acc_online}
    \end{subfigure}
    \hfill
    \begin{subfigure}{.45\textwidth}
        \centering
        \includegraphics[width=\linewidth]{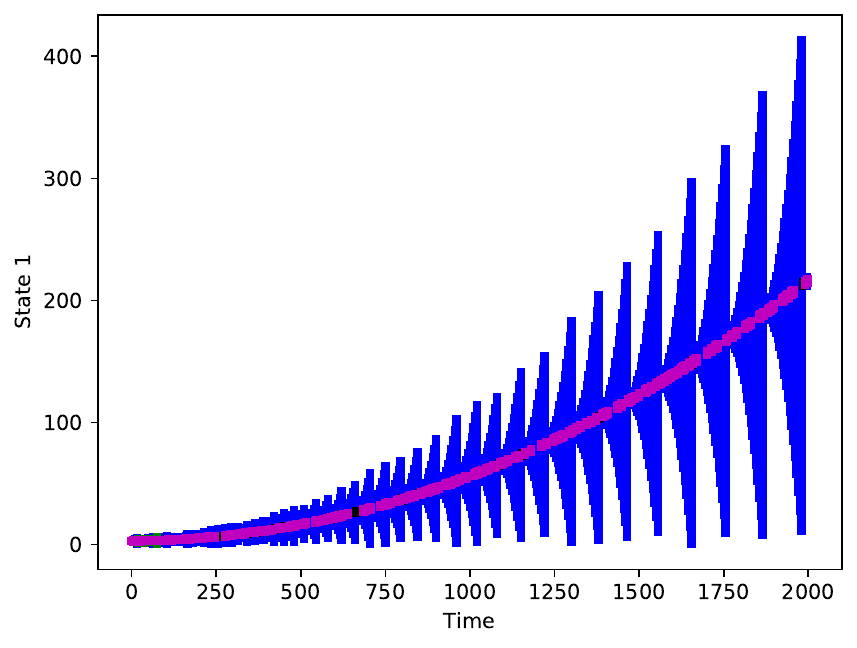}
        \caption{\textbf{Compare Online and Offline Monitoring}}
        \label{fig:acc_comp}
    \end{subfigure}
    \caption{\textbf{\textit{Online Monitoring (ACC).}} We plot the change in distance between two vehicle $h$ with time. The color coding is same as \cref{fig:onlinePKPD}. \textit{Online Monitoring (\cref{fig:acc_online})}: We apply our online monitoring to the ACC model. \textit{Compare (\cref{fig:acc_comp})}: We compare our online and offline algorithms. These plots are a stochastic recreation of~\cite[Figure 6]{GA22}} 
    \label{fig:onlineACC}
\end{figure*}

The results to be recreated for the Anesthesia case study are given in \cref{fig:offlinePKPD,fig:onlinePKPD}.
The results to be recreated for the ACC case study are given in \cref{fig:offlineACC,fig:onlineACC}.

\end{document}